\documentclass[pre,showpacs,preprint]{revtex4}

\usepackage{amsmath}
\usepackage{xcolor}
\usepackage{graphicx}
\usepackage{dcolumn}
\usepackage{bm}

\begin{document}

\title{Stability analysis of the homogeneous hydrodynamics of a model for a confined granular gas}
\author{J. Javier Brey, V. Buz\'{o}n, M.I. Garc\'{\i}a de Soria,  and P. Maynar}
\affiliation{F\'{\i}sica Te\'{o}rica, Universidad de Sevilla,
Apartado de Correos 1065, E-41080, Sevilla, Spain}
\date{\today }

\begin{abstract}
The linear hydrodynamic stability of a model for  confined quasi-two-dimensional granular gases is analyzed. The system exhibits homogeneous
hydrodynamics, i.e. there are macroscopic evolution equations for homogeneous states. The stability analysis is carried out around all these states and not only the homogeneous steady state reached eventually by the system. It is shown that in some cases the linear analysis is not enough to reach a definite conclusion on the 
stability, and molecular dynamics simulation results are presented to elucidate these cases. The analysis shows the relevance of nonlinear hydrodynamic contributions to describe the behavior of spontaneous fluctuations occurring in the system, that lead even to the transitory formation of clusters of particles.
The conclusion is that the system is always stable. The relevance of the results for describing the instabilities of confined granular gases observed experimentally is discussed. 

\end{abstract}

\pacs{45.70.Mg,05.20.Jj,51.10.+y}

\maketitle

\section{Introduction}
\label{s1}
In the last years, a particular geometry has attracted the interest in the study of granular media, 
both experimentally \cite{MVPRKEyU05,RIyS06,RIyS07,RPGRSCyM11,CMyS12} and theoretically \cite{KyA11,BRyS13,SRyB14,KyT15}. It is a quasi-two-dimensional system of spherical particles  placed between two large parallel plates
separated a distance smaller than two particle diameters, so that the particles can not jump on one another.
The container is vertically vibrated to inject energy and maintain the system fluidized. When the system is seen from above, or from below, it looks like
a two-dimensional granular fluid. In another series of experiments, a vibrated monolayer of grains  is also considered, but the system is open on the top \cite{OyU98,GSVyP11,PGGSyV12}.
A peculiarity of these set-ups is that energy is injected in the bulk of the two-dimensional system, instead of at the boundaries, then  allowing the generation of homogeneous reference states. Actually, experiments show that  the horizontal dynamics  remains homogeneous under a large range of values of the parameters defining the system. On the other hand, when increasing the average density and/or decreasing the intensity of the vibration, the system exhibits a series of phase transitions \cite{OyU98,RPGRSCyM11,CMyS12} and, in particular, a bimodal regime  characterized by a single dense cluster of closely grains surrounded by a gas of quite agitated particles \cite{OyU98,CMyS12,KyT15}.  Several models have been proposed trying to describe the effective two-dimensional dynamics of the vibrated system confined between two plates, the main issue being how to incorporate in the dynamics the mechanism of energy injection. A conceptually simple way is to consider an external noise term acting on each particle, so that the inelastic collisions are described in the usual way but the particles  are subjected to random kicks between collisions \cite{WyMc96,vNyE98,PLMyV99,PTvNyE01}. Although it is true that this modeling leads to the existence of a uniform steady state, its possible relation with the state generated in the experiments has not been established. Moreover, this  stochastic driving does not conserve momentum. A very appealing alternative has been proposed recently \cite{BRyS13}. Taking into account that collisions between particles are the mechanism by which the kinetic energy acquired by the particles at the vibrating walls is transferred from the vertical degree of freedom to the horizontal ones,  the idea is to modify the usual two-dimensional collision rule for inelastic disks,  in order to incorporate a description of  the energy transfer. This is done by introducing a characteristic velocity parameter $\Delta$ that is added to each particle in the direction of the normal component of their relative velocity at the collision. Consequently, the normal component of the relative velocity is increased by $2 \Delta$ in each collision, in addition and independently of the effect of inelasticity as described by the coefficient of normal restitution.  Of course, this simple model assumes implicitly that there is no kinetic energy injection in the horizontal directions in the collisions of the particles with the plates, so no friction with the walls is considered. 

Since the model is formulated at the level of particle dynamics, the methods of non-equilibrium statistical mechanics and kinetic theory developed for inelastic hard spheres and disks can be easily extended and applied to the new kinetics \cite{BGMyB13}. In this way, hydrodynamic equations to Navier-Stokes order have been derived for dilute gases described by the Boltzmann-like equation, with identification of the associated transport coefficients, which are given by  the solutions of a system of  first order differential equations \cite{BBMyG15}. A peculiarity of the hydrodynamic equations, as compared with the Navier-Stokes equations of  a normal fluid, is that the time derivatives of the hydrodynamic fields do not vanish in the homogeneous limit, i.e. there is an homogeneous hydrodynamics. This follows from the inherent  non-equilibrium character of granular matter and, actually, it seems to be a  quite general feature of systems exhibiting steady non-equilibrium states \cite{Lu06,GMyT12}. Of course, in the long time limit the homogeneous hydrodynamic equations lead to the steady state of the system.

Upon studying the linear hydrodynamic stability of the system and the possible existence of instabilities driving the system to inhomogeneous configurations, it is important to carry out the analysis around the time dependent state defined by the homogeneous hydrodynamic equations, instead of considering linear deviations from the steady state. The analysis is trivially more general, providing information about the stability not only of the vicinity of the steady state but also on the homogeneous hydrodynamic trajectory relaxing towards it, i.e. hydrodynamic states outside the linear environment of the steady state are included in the analysis. 
It is possible that a system be stable near the homogeneous steady state but it be unstable for homogeneous configurations corresponding to values of the macroscopic fields far away from the steady ones.

In this paper, the linear hydrodynamic stability of the  homogeneous evolution of the quasi-two-dimensional granular model is studied. The aim is to see whether the model is able to predict some kind of hydrodynamic instability of the homogeneous state and, hence, to open the possibility of some phase coexistence similar to the one observed in the experiments. Moreover, the analysis presented serves to illustrate how the stability of general steady non-equilibrium states of molecular systems must be addressed \cite{Lu06}. It will be shown that linear perturbations around some homogeneous states grow in the short time limit, although the linearized hydrodynamic equations predict that they decay afterwards.  It follows that for these states a definite answer regarding its stability can not be reached from the linear analysis. Consequently, molecular dynamics simulations of some of these states have been performed. They indicate that the system actually relaxes to the homogeneous time-dependent state, but that the mechanism seems to be quite complex, involving highly nonlinear hydrodynamic effects.

 The remaining of the paper is organized as follows. In Sec. \ref{s2}, the linear hydrodynamic equations around the time-dependent hydrodynamic homogeneous state are presented. They are written using dimensionless perturbations of the hydrodynamic fields around homogeneity, and have the form of a system of coupled first order diferencial equations with time-dependent coefficients. Its solution is rather involved and has to be done numerically, except for the special case of the traverse velocity field, whose equation is decoupled from the rest and its solution can  be written explicitly in terms of a well defined shear mode.  The linear stability analysis is presented in Sec. \ref{s3}. The results imply the need of a nonlinear analysis of the stability for some homogeneous states, something that seems to be too complicated. Therefore,    molecular dynamics simulations regarding these states have been performed and the results are shown in Sec. \ref{s4}. The simulation results indicate that the homogeneous time-dependent states are all stable, although relevant inhomogeneous fluctuations can be observed on short time scales. The last section of the paper contains a short summary of the results and some comments on their relevance to explain the inhomogeneous states observed experimentally.

\section{The linear Navier-Stokes equations around homogeneous hydrodynamics}
\label{s2}
The system considered is a granular gas composed of smooth inelastic spheres of mass $m$ and diameter $\sigma$, confined to a quasi-two dimensional geometry by means
of two large parallel horizontal plates separated a distance smaller than two particle diameters. Energy is injected into the system 
by vibrating the parallel walls, and the rate of energy injection is large enough as to render negligible the effect of the gravitational potential energy.  Interest is focussed on the two dimensional dynamics showed by the system when observed from above. 
This dynamics has been assimilated to that of a two-dimensional granular fluid, and a kinetic model has been proposed to describe the effective 
dynamics in the plane \cite{BRyS13}. The projections of the spheres are considered as inelastic hard disks of mass $m$ and diameter $\sigma$.
To incorporate a mechanism trying to describe  the transferring of the energy injected vertically to the horizontal dynamics, the usual collision rule of smooth inelastic hard disks is modified, so that when two disks with velocities ${\bm v}_{1}$ and ${\bm v}_{2}$ collide, their velocities are instantaneously changed to new values
given by 
\begin{equation}
\label{2.1}
{\bm v}^{\prime}_{1} = {\bm v}_{1}- \frac{1+ \alpha}{2}  {\bm v}_{12} \cdot \widehat{\bm \sigma}  \widehat{\bm \sigma} + \Delta \widehat{\bm \sigma},
\end{equation}
\begin{equation}
\label{2.2}
 {\bm v}^{\prime}_{2} = {\bm v}_{2}+ \frac{1+ \alpha}{2}  {\bm v}_{12} \cdot \widehat{\bm \sigma} \widehat{\bm \sigma} - \Delta \widehat{\bm \sigma},
\end{equation}
where $\widehat{\bm \sigma}$ is the unit vector joining the centers of the two disks at contact, ${\bm v}_{12} \equiv {\bm v}_{1}- {\bm v}_{2}$ is the relative velocity,
$\alpha$ is the coefficient of normal restitution defined in the interval $ 0 < \alpha \leq 1$,  and $\Delta$ is some positive characteristic speed. This is the quantity  trying to describe the way in which the  energy goes from the vertical degree of freedom to the horizontal ones in the collisions of the original hard spheres. Note that the above collision rule conserves momentum in the plane.
Since the model is formulated at the collisional level, the methods of nonequilibrium statistical mechanics and kinetic theory developed for inelastic hard spheres and disks \cite{BDyS97,vNEyB98} can be easily adapted \cite{BGMyB13}. From the Liouville equation, balance equations for the macroscopic fields, the number of particles density, $n({\bm r},t)$, the velocity flow, ${\bm u}({\bm r},t)$, and the granular temperature, $T({\bm r},t)$, are derived in a straightforward way,
\begin{equation}
\label{2.3}
\frac{\partial n}{\partial t} +  {\bm \nabla} \cdot \left( n {\bm u} \right) =0,
\end{equation}
\begin{equation}
\label{2.4}
\frac{\partial {\bm u}}{\partial t} +{\bm u} \cdot {\bm \nabla} {\bm u} + (mn)^{-1} {\bm \nabla}
\cdot \sf{P}=0,
\end{equation}
\begin{equation}
\label{2.5}
\frac{\partial T}{\partial t} +{\bm u} \cdot {\bm \nabla}T + \frac{1}{n} \left( {\sf P} : {\bm \nabla} {\bm u} + {\bm \nabla} \cdot {\bm J}_{q} \right) = - \zeta T.
\end{equation}
In the above equations, ${\sf P}({\bm r},t) $ is the pressure tensor, ${\bm J}_{q}({\bm r},t)$ is the heat flux, and $\zeta ({\bm r}, t)$ is the rate of change of the temperature due to the inelasticity of collisions. These three quantities are obtained as functionals of the distribution function of the system. In order to close the balance equations, it is necessary to express  ${\sf P}$, ${\bf J}_{q}$ and $\zeta$ in terms of the macroscopic  fields by means of some constitutive relations. In the low density limit, the
dynamics of the system is expected to be accurately described by the Boltzmann equation with the same degree of confidence as for elastic hard disks \cite{BGMyB13}. Then, by using an extension of the Chapman-Enskog procedure, it is possible to derive expressions for ${\sf P}$, ${\bf J}_{q}$, and 
$\zeta$, in the form of series expansions in the gradients of the macroscopic fields. To first order, they have the form \cite{BBMyG15},
\begin{equation}
\label{2.6}
{\sf P}= n T {\sf I}- \eta \left[ {\bm \nabla} {\bm u} + \left( {\bm \nabla} {\bm u} \right)^{+} - {\bm \nabla} \cdot {\bm u} {\sf I} \right],
\end{equation}
\begin{equation}
\label{2.7}
{\bm J}_{q}= - \kappa {\bm \nabla} T - \mu {\bm \nabla}  n,
\end{equation}
\begin{equation}
\label{2.8}
\zeta= \zeta^{(0)}+ \zeta_{1} {\bm \nabla} \cdot {\bm u}.
\end{equation}
In Eq. (\ref{2.6}),  ${\sf I}$ is the two-dimensional unit tensor, $\left( {\bm \nabla} {\bm u} \right)^{+} $ is the transposition of ${\bm \nabla} {\bm u} $, and $\eta$
is the coefficient of shear viscosity. In Eq. (\ref{2.7}), $\kappa$ is the coefficient of (thermal) heat conductivity and $\mu$ is the coefficient of diffusive heat conductivity. The latter is peculiar of systems with inelastic collisions \cite{BMyD96}, vanishing in the elastic limit. Finally, the expression of the rate of change of the temperature contains a zeroth order in the gradients term, $\zeta^{(0)}$, and an Euler transport coefficient $\zeta_{1}$, which is also peculiar of inelastic collisions \cite{GyD99,BDyB08,DByB08}. For non-confined granular gases, it vanishes in the low density limit.  In principle, the second order in the gradients contributions to the rate of change of the temperature $\zeta$ should be also considered, since they lead to terms in Eq. (\ref{2.5}) which are of the same order than the contributions coming from Eqs. (\ref{2.6}) and (\ref{2.7}).  Nevertheless, these new terms are expected to be quantitatively negligible as compared with those terms from the heat flux and the pressure tensor. The expression of $\zeta^{(0)}$, as well as the differential equations determining all the other transport coefficients were derived  in Ref. \cite{BBMyG15}, and for the sake of completeness they are reproduced in the Appendix. The general structure of the hydrodynamic equations can be justified by means of rather general arguments \cite{BRyS13}. Nevertheless, it is important to realize that the presence of the characteristic velocity $\Delta$  implies that a nondimensional parameter can be constructed as $\Delta^{*} \equiv \Delta / v_{0}(T)$, with $v_{0}(T) \equiv \left( 2T/m \right)^{1/2}$ being a thermal velocity. As a consequence, the dependence  of the transport coefficients on the temperature is much more involved in the present case than in systems of elastic or inelastic hard disks.

Substitution of Eqs. (\ref{2.6})-(\ref{2.8}) into Eqs. ({\ref{2.3})-(\ref{2.5}) gives the Navier-Stokes equations describing the hydrodynamics of the system. For homogeneous situations, they reduce to
\begin{equation}
\label{2.9}
\frac{\partial n_{H}}{\partial t}=0, 
\end{equation}
\begin{equation}
\label{2.10}
\frac{\partial {\bf u}_{H}}{\partial t}=0,
\end{equation}
\begin{equation}
\label{2.11}
\frac{\partial T_{H}}{\partial t} = -\zeta_{H}^{(0)} T_{H}.
\end{equation}
Here, and in the following, the subindex $H$ is used to denote a quantity computed in an hydrodynamic homogeneous state. Then, e.g. $\zeta_{H}^{(0)} \equiv \zeta^{(0)}(n_{H},T_{H})$,  and so on. Without lost of generality, the value ${\bm u}_{H}=0$ can be taken.  The above equations predict the existence of an homogeneous steady state with a temperature $T_{st}$ determined by the equation $\zeta^{(0)}(n_{H}, T_{st})=0$. This state has been extensively studied and the theoretical predictions from hydrodynamics have been shown to be in good agreement with molecular dynamics simulation results \cite{BRyS13, BGMyB13,SRyB14}, The aim here is to investigate the stability of the homogeneous hydrodynamic states, i.e. to determine whether the evolution of a homogeneous system as described by the above hydrodynamic equations is linearly stable. The particular issue of the stability of the steady homogeneous state has already been addressed, with the results that hydrodynamics predicts that it is linearly stable \cite{BRyS13}. To start with, dimensionless deviations of the macroscopic fields from homogeneity are defined as
\begin{equation}
\label{2.12}
\rho({\bm r},t) \equiv \frac{n({\bm r},t)-n_{H}}{n_{H}},
\end{equation}
\begin{equation}
\label{2.13}
{\bm \omega} ({\bm r},t) \equiv \frac{{\bm u} ({\bm r},t)}{v_{0H}(t)},
\end{equation}
\begin{equation}
\label{2.14}
\theta ({\bm r},t) \equiv \frac{T({\bm r},t)-T_{H}(t)}{T_{H}(t)}.
\end{equation}
Moreover, it is convenient to use also dimensionless time scale $s$ and lenght scale ${\bm l}$ defined by
\begin{equation}
\label{2.15}
ds \equiv \frac{v_{0H}(t)}{\ell}\, dt
\end{equation}
and
\begin{equation}
\label{2.16}
d{\bm l} \equiv \frac{d {\bm r}}{\ell},
\end{equation}
respectively. The unit of length, $\ell \equiv (n_{H}  \sigma )^{-1}$ is proportional to the mean free path of the particles and the time scale $s$ is proportional to the cumulative number of collisions per particle in the associated original time interval, both quantities computed in the time dependent reference homogeneous state. 

The linearized equations will be written in the Fourier representation, with the transforms of the hydrodynamic fields defined as
\begin{equation}
\label{2.17}
\widetilde{\rho}_{\bm k}(t)  \equiv \int d{\bm l}\, e^{-i {\bm k} \cdot {\bm l}} \rho ({\bm l},t),
\end{equation}
and so on.  After some straightforward algebra, linearization of Navier-Stokes equations in $\rho$, ${\bm \omega}$, and $\theta$ leads to 
\begin{equation}
\label{2.18}
\frac{\partial \widetilde{\rho}_{\bm k}}{\partial s} + i {\bm k} \cdot \widetilde{\bm \omega}_{\bm k}=0,
\end{equation}
\begin{equation}
\label{2.19}
\frac{\partial \widetilde{\bm \omega}_{\bm k}}{\partial s} - \frac{\overline{\zeta}_{H}^{(0)}}{2}\,  \widetilde{\bm \omega}_{\bm k} + \frac{i{\bm k}}{2} \left( \widetilde{\theta}_{\bm k} + \widetilde{\rho}_{\bm k} \right) + \overline{\eta}_{H} \widetilde{\bm \omega}_{\bm k} k^{2}=0,
\end{equation}
\begin{equation}
\label{2.20}
\frac{\partial \widetilde{\theta}_{\bm k}}{\partial s}+(1+\zeta_{1H})i {\bm k}\cdot  \widetilde{\bm \omega}_{\bm k} + k^{2} \left( \overline{\kappa}_{H} \widetilde{\theta}_{\bm k} + \overline{\mu}_{H} \widetilde{\rho}_{\bm k} \right)=- \widetilde{\rho}_{\bm k} \overline{\zeta}_{H}^{(0)}- \widetilde{\theta}_{\bm k} \psi_{H}, 
\end{equation}
where dimensionless coefficients have been defined as
\begin{equation}
\label{2.21}
\overline{\zeta}^{(0)}_{H}  \equiv \frac{\zeta_{H}^{(0)} \ell }{ v_{0H}}\, ,
\end{equation}
\begin{equation}
\label{2.22}
\overline{\eta}_{H} \equiv \frac{\eta_{H} }{m n_{H}\ell v_{0H}}\, ,
\end{equation}
\begin{equation}
\label{2.23}
\overline{\kappa}_{H} \equiv \frac{\kappa_{H}}{n_{H} \ell v_{0H}}\, ,
\end{equation}
\begin{equation}
\label{2.24}
\overline{\mu}_{H}  \equiv \frac{\mu_{H}}{T_{H} \ell v_{0H}}\, .
\end{equation}
The term $\psi$ in the equation for the temperature deviation $\widetilde{\theta}_{\bm k}$ is given by
\begin{equation}
\label{2.25}
\psi ( \Delta^{*}) =
\left( \frac{\pi}{2} \right)^{1/2} \left\{ 
\frac{1-\alpha^{2}}{2} \left( 1 + \frac{3a_{2}}{16} \right) + \left( 1- \frac{a_{2}}{16} \right) \Delta^{*2}- \left[ \frac{3}{32} (1-\alpha^{2}) + \frac{\Delta^{*2}}{16} \right]\frac{\partial a_{2}}{\partial \Delta^{*}} \Delta^{*}  \right\},
\end{equation} 
with $a_{2}$ being the coefficient of the first Sonine correction for the distribution function of homogeneous hydrodynamic states \cite{BMGyB14}}. Its expression has been investigated in ref. \cite{BMGyB14}, showing that it is given by the solution of Eq. (\ref{ap.9}) in the Appendix. As an example, in Fig.\ \ref{fig1} it  is plotted as a function of $\Delta^{*}$ for $\alpha=0.8$.  The steady value of $\Delta^{*}$, $\Delta^{*}_{st}  \simeq  0.15$, is indicated. Since the theory has been developed by assuming $a_{2} \ll1$, the interval of values of $\Delta^{*}$ for which the theory can be expected to hold may be estimated to be bounded  $\Delta^{*} \lesssim 0.5$. This means that when $\Delta$ is large as compared with the thermal velocity $v_{0}(T)$, the first Sonine approximation, and therefore the Gaussian function, fail to describe the distribution function of the system. On the other hand, for large temperatures such that $\Delta^{*} \ll 1$, the state of the system approaches the homogeneous cooling state for which the Sonine approximation is known to provide a fair description of the velocity distribution. The function $\psi ( \Delta^{*})$ is shown in Fig.\ \ref{fig2}.  It is seen to be an increasing function over the relevant interval of the scaled velocity parameter. Similar results are obtained for other values of $\alpha$.

\begin{figure}
\includegraphics[width=.5\textwidth]{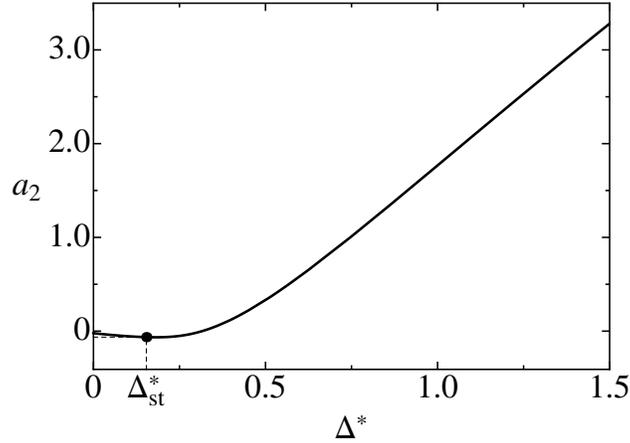}
  \caption{Sonine coefficient $a_{2}$ of the hydrodynamic homogeneous one-particle distribution  as a function of the scaled dimensionless velocity parameter $\Delta^{*}$ for $\alpha = 0.8$. The steady value of $\Delta^{*}$, denoted by $\Delta^{*}_{st}$  is indicated. As discussed in the main text, the first Sonine approximation employed in the theoretical analysis is expected to fail when $a_{2}$ becomes of the order of unity.}
  \label{fig1}
\end{figure}

\begin{figure}
\includegraphics[width=.5\textwidth]{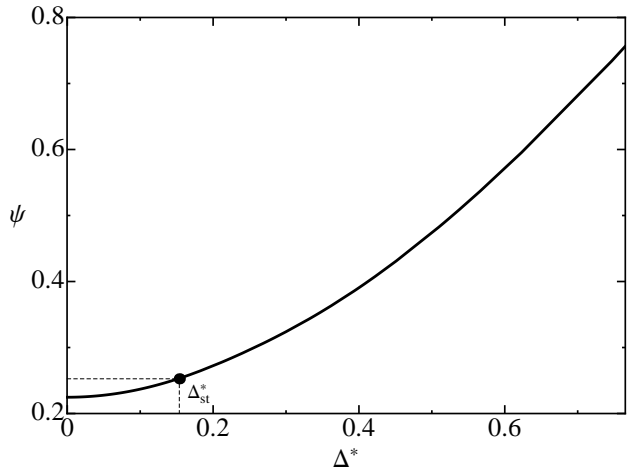}
  \caption{The dimensionless quantity $\psi$, defined by Eq. (\protect{\ref{2.25}}), as a function of the scaled dimensionless velocity parameter $\Delta^{*}$ for a system with $\alpha=0.8$. Only the interval of the velocity in which the Sonine coefficient  $a_{2}$ remains small, say $a_{2} \lesssim 0.5 $ has been plotted. In this relevant region, $\psi$ is  a positive increasing function of $\Delta^{*}$.}
  \label{fig2}
\end{figure}

The equation for the velocity, Eq. (\ref{2.19}), can be decomposed into two equations, for the longitudinal and transversal components of the velocity field relative to the wave vector ${\bm k}$, $\widetilde{\omega}_{{\bm k}, \parallel}$ and $\widetilde{\omega}_{{\bm k}, \perp}$, respectively,
\begin{equation}
\label{2.26}
\frac{\partial \widetilde{\omega}_{{\bm k} \perp}}{\partial s} - \frac{\overline{\zeta}_{H}^{(0)}}{2}\,  \widetilde{\omega}_{\bm k \perp} + \overline{\eta}_{H} \widetilde{\omega}_{\bm k \perp} k^{2}=0,
\end{equation}
\begin{equation}
\label{2.27}
\frac{\partial \widetilde{\omega}_{{\bm k} \parallel}}{\partial s} - \frac{\overline{\zeta}_{H}^{(0)}}{2}\,  \widetilde{\omega}_{\bm k \parallel}+ \frac{i k}{2} \left( \widetilde{\theta}_{\bm k} + \widetilde{\rho}_{\bm k} \right) + \overline{\eta}_{H} \widetilde{\omega}_{\bm k \parallel} k^{2}=0.
\end{equation}
It is worth to stress that, in spite of the change to the number of collisions time scale $s$, the coefficients of the linearized hydrodynamic equations still depend on time, contrary to what happens in systems of smooth inelastic hard spheres or disks when considering the linearization around the homogeneous cooling state \cite{BDKyS98}. On the other hand, linear hydrodynamic equations with coefficients having a nonsimple time dependence have been also found in a model for balistic annihilation \cite{GMSByT08} and a granular gas in contact with a thermostat \cite{GFHyY16}

\section{Stability analysis} 
\label{s3}

As pointed out above, the coefficients of the linearized hydrodynamic equations depend on time in rather  different ways due to the presence of the velocity parameter $\Delta$ in the formulation of the dynamics of the system. There is a  time dependence that occurs through different powers of the dimensionless parameter $\Delta^{*} \propto  \Delta / \sqrt{T(t)}$ and that, consequently, can not be scaled out in a direct way. Therefore, the complete linear stability analysis becomes more difficult than for smooth inelastic hard spheres in the homogeneous cooling state, since both the eigenvalues and eigenfunctions of the dimensionless problem are time-dependent. An exception is the particular case of the steady state, since it is time-independent by definition. For this state, the stability analysis is straightforward \cite{BRyS13}, and shows that the steady state is linearly stable. In this context, it is worth to notice that the steady state reached when a (non-confined) granular gas is thermalized by means of the so-called stochastic thermostat, i..e by considering a white noise force acting on each particle, is also linearly stable \cite{GMyT13,GCyV13}.

For a general homogeneous state of the confined granular gas, the linearized equation  for the transversal velocity field $\widetilde{\bm \omega}_{{\bm k} \perp}$, Eq. (\ref{2.26}),  is decoupled from the rest and can be integrated directly yielding
\begin{equation}
\label{3.1}
\widetilde{\bm \omega}_{{\bm k} \perp} (s)= \widetilde{\bm \omega}_{{\bm k} \perp}(0) \exp \int_{0}^{s} d s^{\prime}\, \lambda_{\perp}(s^{\prime})\, ,
\end{equation}
where
\begin{equation}
\label{3.2}
\lambda_{\perp}(s)  \equiv \frac{\overline{\zeta}^{(0)}}{2} - \overline{\eta}_{H} k^{2}.
\end{equation}
This identifies the shear mode. In order to analyze its behavior,  it is convenient to differentiate between the two parameter regions: $\Delta^{*}_{H}> \Delta^{*}_{st}$ and $\Delta^{*}_{H}< \Delta^{*}_{st}$. In the former, the time-dependent temperature of the reference homogeneous state is smaller than its stationary value, and $\Delta^{*}_{H}$ is a decreasing function of time, i.e. the homogeneous reference state is heating and, because of Eq.\ (\ref{2.11}),  the rate of change of the homogeneous temperature is negative, $\overline{\zeta}_{H}^{(0)} < 0$, and hence also  $\lambda_{\perp}(s) <0$, since the shear viscosity coefficient is always positive. The conclusion is that the scaled shear mode is  linearly stable for those homogeneous hydrodynamic processes  in which the system heats towards the steady state. On the other hand, when $\Delta^{*}_{H}< \Delta^{*}_{st}$, the system is cooling and $\overline{\zeta}_{H}^{(0)} > 0$. Therefore, $\lambda_{\perp}(s)$ is positive if $k<k_{\perp}(s)$ with
\begin{equation}
\label{3.3}
k_{\perp}(s)= \left(  \frac{\overline{\zeta}_{H}^{(0)}}{2 \overline{\eta}_{H}} \right)^{1/2} .
\end{equation}
This value of the wavenumber dependes on time through $\Delta^{*}_{H}$. Actually, since the cooling rate tends to zero as the stationary state is approached, while the shear viscosity remains finite and positive, it follows that $k_{\perp}$ vanish in the long time limit. The special case of an homogeneous ($k=0$) perturbation of the transversal velocity field deserves a comment. It is trivially seen that $\widetilde{\bm \omega}_{0 \perp} $ grows monotonically in time until the steady state is reached, but this behavior is a direct consequence of the scaling of the velocity field  with the square root of the temperature, Eq. (\ref{2.13}). The actual, unscaled velocity field remains constant, with the same initial value of the perturbation.

\begin{figure}
\includegraphics[width=.7\textwidth]{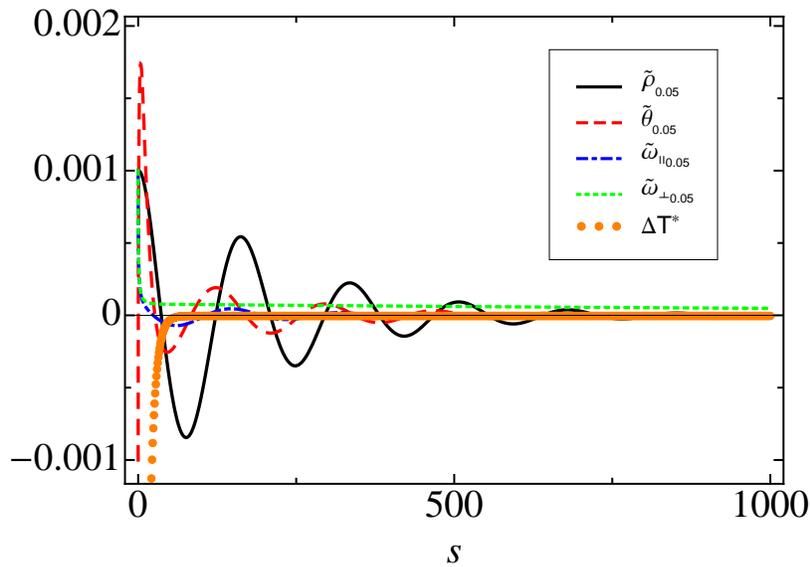}
  \caption{(Color online) Time evolution of the perturbations of the dimensionless hydrodynamic fields in a confined quasi-two-dimensional granular gas with $\alpha=0.9$ as predicted by the linearized hydrodynamic equations. Time is measured in the dimensionless scale $s$ defined in the main text and the wavenumber is $k=0.05$.  The initial temperature of the system is smaller than its stationary value, so that the reference temperature  is monotonically increasing in time. The initial values of the perturbations are $\widetilde{\theta}_{0.05}=-10^{-3} $ and $\widetilde{\rho}_{0.05}=\widetilde{\omega}_{0.05 \perp }= \widetilde{\omega}_{0.05 \parallel}=10^{-3}$.}
  \label{fig3}
\end{figure}

With regards to the other macroscopic fields, although quite involved, it is possible to numerically integrate the coupled linearized hydrodynamic equations
(\ref{2.18}), (\ref{2.20}), and (\ref{2.27}). Figure \ref{fig3} displays the evolution of all the linearized hydrodynamic fields as a function of the number of collisions scale $s$  in a system with a coefficient of normal restitution $\alpha=0.9$ and an initial temperature smaller than the steady one, namely $\Delta^{*}(0) =1> \Delta^{*}_{st} \simeq 0.078$. The wavenumber considered  is $k= 0.05$.   Also plotted for reference of the time scales is the evolution of the temperature of the homogeneous state measured by the parameter $\Delta T^{*}(s) \equiv (T(s)-T_{st})/100 T_{st}$.

The initial perturbations have quite small amplitudes, as required for the validity of the linear hydrodynamic equations. Their values are given in the figure caption. It is observed that the temperature, density, and velocity  fields oscillate in time indicating a behavior that is peculiar of complex hydrodynamic modes, like the elastic  sound modes. Also, the amplitude of the oscillations decays,  i.e. the initial inhomogeneities tend to vanish, following that  the system is linearly stable. 

\begin{figure}
\includegraphics[width=.7\textwidth]{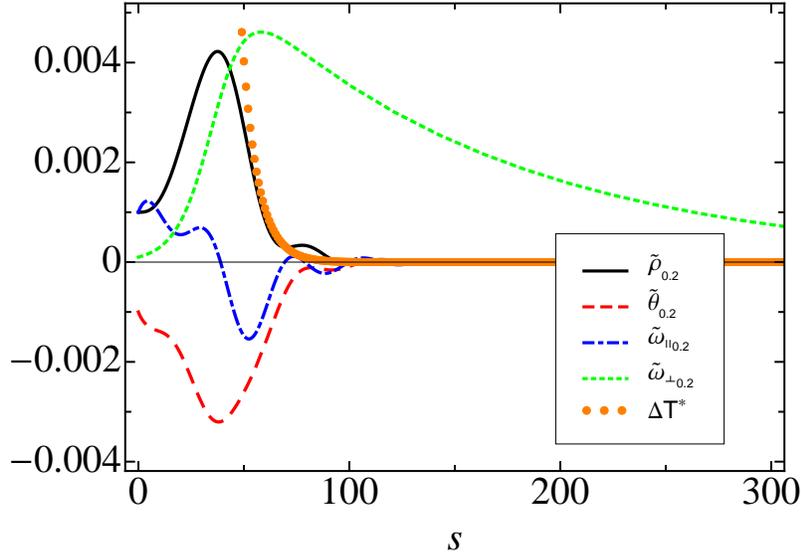}
  \caption{(Color online) Time evolution of the perturbations of the dimensionless hydrodynamic fields in a confined quasi-two-dimensional granular gas  with $\alpha=0.9$ as predicted by the linearized hydrodynamic equations. Time is measured in the dimensionless scale $s$ defined in the main text and the wavenumber is $k=0.2$.  The initial temperature of the system is larger than its stationary value so that the reference temperature  is monotonically decreasing  in time. The initial values of the perturbations  are $\widetilde{\theta}_{0.2}=-10^{-3} $, $\widetilde{\rho}_{0.2}=  \widetilde{\omega}_{0.2 \parallel}=10^{-3}$, and $\widetilde{\omega}_{0.2 \perp }= 10^{-4}$.}
  \label{fig4}
\end{figure}

 The behavior of the perturbations in a system evolving with a temperature larger than the steady value is illustrated in Fig. \ref{fig4}, again for a system with $º\alpha=0.9$, being now $\Delta^{*}(0) = 10^{-3}$.  The initial value of $k_{\perp}$ defined in Eq.\ (\ref{3.3}) is $k_{\perp} \simeq 0.757$, and the value of the wavenumber used in the simulations is $k= 0.2$ . Therefore, the perturbation of the scaled transversal velocity, $\widetilde{\omega}_{0.2 \perp}$, is predicted to initially grow on time by its linear hydrodynamic equation, as it is observed in the figure. In fact, the amplitudes of the perturbations of the other fields also increase, although a superimposed oscillatory behavior is identified. Nevertheless, in the long time limit all the hydrodynamic perturbations tend to vanish, with the transversal velocity decaying rather slower than the other fields. Similar behaviors have been found for other values of the initial perturbations and the wavenumber $k$. The picture emerging from the linear analysis is that although the traversal modes can grow for a transitory period, the speed of growth is smaller that the rate of change of the eigenvalue, so that the latter change from positive to negative before the mode grows too much. Then, one is tempted to conclude that the homogeneous hydrodynamic states are linearly stable, but it must be  taken into account that if the amplitude of the perturbation grows     enough as to leave the linear regime, the description provided by Eqs. (\ref{2.18})-(\ref{2.20}) breaks down, making it necessary to consider the complete nonlinear hydrodynamic equations.

 \section{Molecular dynamics simulations}
 \label{s4}       
         
 An important issue raised in the previous section is  whether the amplitude of  a small perturbation of the hydrodynamic fields can   grow enough as to leave the range in which the description provided by the linearized hydrodynamic equations  is valid, so that a more complete non-linear mode description would be required. If this is the case, a second question is whether the  system finally returns to the homogeneous time-dependent hydrodynamic state, so that this state is stable. To investigate the above subjects, molecular dynamics (MD) simulations of a system of N  inelastic hard disks in  a square of size $L$, obeying the modified collision rules given by Eqs. (\ref{2.1}) and (\ref{2.2})  have been performed. An event-driven algorithm \cite{AyT87} with periodic boundary conditions has been used. The number density in all the simulations has been $ n= 0.05 \sigma^{-2}$, that is low enough as to consider that the system is in the low density limit, at least as long as it does not have too large density inhomogeneities. Initially, the particles were distributed uniformely into the system and their velocities obeyed a Gaussian distribution. In the simulations to be reported  it is $N=2000$, and the initial granular temperature $T(0)$ was much larger than its steady value, namely $T_{st}= 9,7\times 10^{-8} T(0)$, in such a way that the homogeneous reference state is cooling towards the steady state, i.e. the kind of processes for which the linear stability analysis does not predict the monotonic decay of hydrodynamic perturbations.
 A relevant feature to keep in mind when analyzing the simulation results is that the observed perturbations are spontaneous, i.e. generated by the the dynamics of the system itself, and not externally introduced. Consistently, the data displayed correspond to a single computer run or phase space trajectory of the system, since average over different trajectories would smooth down the spontaneous fluctuations.
 
 The hydrodynamic fields measured in the simulations are defined as
 \begin{equation}
 \label{4.1}
 \widetilde{u}_{x,\perp, k_{min}}^{(1)}=  \sum_{i=1}^{N} \frac{v_{i,x}}{v_{0}(t)} \cos \frac{2 \pi y_{i}}{L}\, ,
 \end{equation}
 \begin{equation}
 \label{4.2}
  \widetilde{u}_{x,\perp, k_{min}}^{(2)}=  \sum_{i=1}^{N} \frac{v_{i,x}}{v_{0}(t)} \sin \frac{2 \pi y_{i}}{L}\, ,
 \end{equation}
 \begin{equation}
 \label{4.3}
 \widetilde{u}_{x,\parallel, k_{min}}^{(1)}=  \sum_{i=1}^{N} \frac{v_{i,x}}{v_{0}(t)} \cos \frac{2 \pi x_{i}}{L}\, ,
 \end{equation}
  \begin{equation}
  \label{4.4}
   \widetilde{u}_{x,\parallel, k_{min}}^{(2)}= \sum_{i=1}^{N} \frac{v_{i,x}}{v_{0}(t)} \sin \frac{2 \pi x_{i}}{L}\, ,
 \end{equation}   
\begin{equation}
\label{4.5}
\widetilde{n}_{k_{min}}^{(x,1)}=  \sum_{i=1}^{N} \cos \frac{2 \pi x_{i}}{L}\, ,
\end{equation}
\begin{equation}
\label{4.6}
\widetilde{n}_{k_{min}}^{(x, 2)}= \sum_{i=1}^{N} \sin \frac{2 \pi x_{i}}{L}\, .
\end{equation}
The expressions correspond to Fourier components with a wave vector $k_{min} \equiv \frac{2 \pi \ell}{L}$  that is the minimum wavenumber compatible with the applied  periodic boundary conditions.  This choice of $k=k_{min}$ corresponds to the largest value of the eigenvalue $\lambda_{\perp}$ associated to the shear mode in the short time limit [See Eq. (\ref{3.2})] and, therefore, it is the one expected to show more clearly an unstable behavior, if it is exhibited by the system. A second reason for this choice is that the used hydrodynamic equations hold in the limit of small gradients of the macroscopic fields or, equivalently, small wavevectors.

Equations (\ref{4.1}) and (\ref{4.2}) define the two Fourier components of the instantaneous  transversal velocity field for the wavevector  ${\bm k}$ in the direction of the $y$-axis.  The Fourier components of the longitudinal velocity field for wavevector in the direction of the $x$-axis  are defined by Eqs. (\ref{4.3}) and (\ref{4.4}). Finally, Eqs.\ (\ref{4.5}) and (\ref{4.6}) are the definitions of the two Fourier components of the number density field
 for a wavevector $k=k_{min}$ along the $x$ direction. Of course, analogous definitions can be made by interchanging the $x$ and $y$ components of both ${\bm r}_{i}$ and ${\bm v}_{i}$. They should lead to equivalent results, by symmetry considerations.
 
 Figure \ref{fig5} displays the evolution of the Fourier components of the velocity field as a function of the accumulated number of collisions per particle $\tau$, which is proportional to the  time scale $s$ used in the previous sections. The coefficient of normal restitution is $\alpha=0.7$ and the characteristic speed is $\Delta = 10^{-4} v_{0}(0)/\sqrt{2}$. It is observed that an spontaneous fluctuation of the transversal velocity field occurs and that its deviation from the steady vanishing value grows quite fast reaching a maximum for $\tau \simeq 50$. Afterwards, the fluctuation decays becoming imperceptible for $\tau \gtrsim 150$. On the other hand, the longitudinal velocity components show an oscillatory behavior for all times, although  the amplitude of the oscillations has a maximum at the same time that the transversal velocity fluctuation. Since in the linear regime the time evolutions of the longitudinal and the transversal components of the velocity  are decoupled, it follows that the observed effect is clearly due to some nonlinear coupling between hydrodynamic modes. 
 
The time evolution of the Fourier components of the density field for the same system and  phase space trajectory is shown in Fig.\ \ref{fig6}. All the components exhibit oscillations around the homogeneous vanishing value. The amplitudes  of the oscillations show a well defined maximum for roughly the same time at which the transversal velocity fluctuation has also a maximum in Fig.\ \ref{fig5}. This reflects that the observed density fluctuations are also induced by the fluctuations of the transversal velocity through some nonlinear coupling. This feature is similar to the one happening in the clustering instability of a non-confined granular gas in the homogeneous cooling state \cite{BRyC99}. Actually, in the present case the system also exhibits  transitory clustering effects that are clearly noted by looking at a snapshot of the evolution of the system. An example is given in Fig.\ \ref{fig7}, that corresponds to the  same trajectory as Figs. \ref{fig5} and \ref{fig6}.

 \begin{figure}
\includegraphics[width=.7\textwidth]{bbgym16af5.eps}
  \caption{(Color online)  MD simulation results for time evolution of the dimensionless transversal and longitudinal  components of an spontaneous fluctuation of the $x$ component of the velocity field. The wavenumber $k=k_{min}$ is the minimum compatible with the imposed periodic boundary conditions.  Time $\tau$ is the accumulated number of collisions per particle. The parameters of the system are $\alpha =0.7$, $n=0.05 \times \sigma^{-2}$, $N=2000$, and $\Delta= 10^{-4} v_{0}(0)/\sqrt{2}$, where $v_{0}(0) \equiv \left( 2T(0)/m\right)^{1/2}$. }
  \label{fig5}
\end{figure}

 \begin{figure}
\includegraphics[width=.7\textwidth]{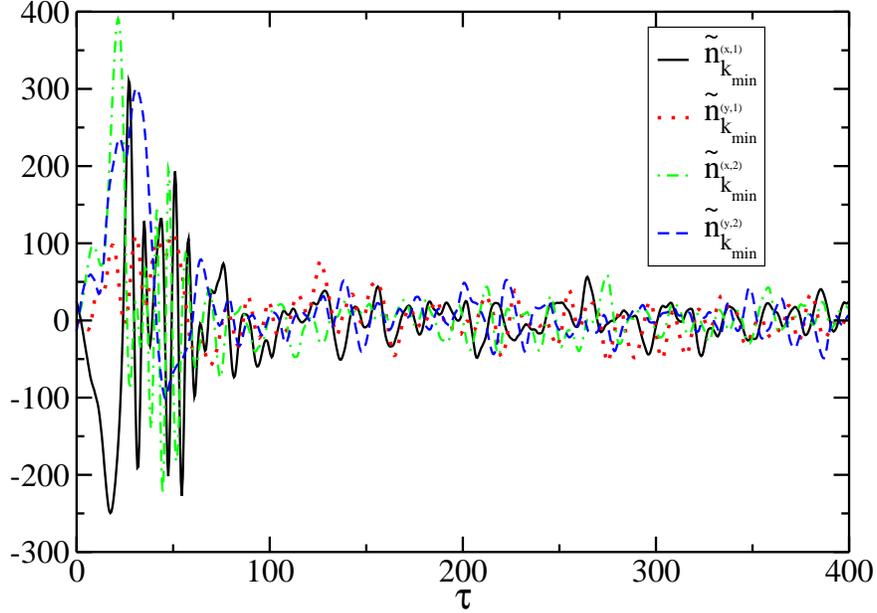}
  \caption{(Color online) MD simulation results for the time evolution of an spontaneous fluctuation of the dimensionless Fourier components of the density field.  The wavenumber $k=k_{min}$ is the minimum compatible with the imposed periodic boundary conditions.  Time $\tau$ is the accumulated number of collisions per particle. The results have been obtained for the same system and from the same phase space trajectory as considered in Fig. \protect{\ref{fig5}}. Note that the two Fourier components corresponding to the wavevector along each of the two  $x$ and $y$ directions are plotted.}
  \label{fig6}
\end{figure}

\begin{figure}
\includegraphics[width=.7\textwidth]{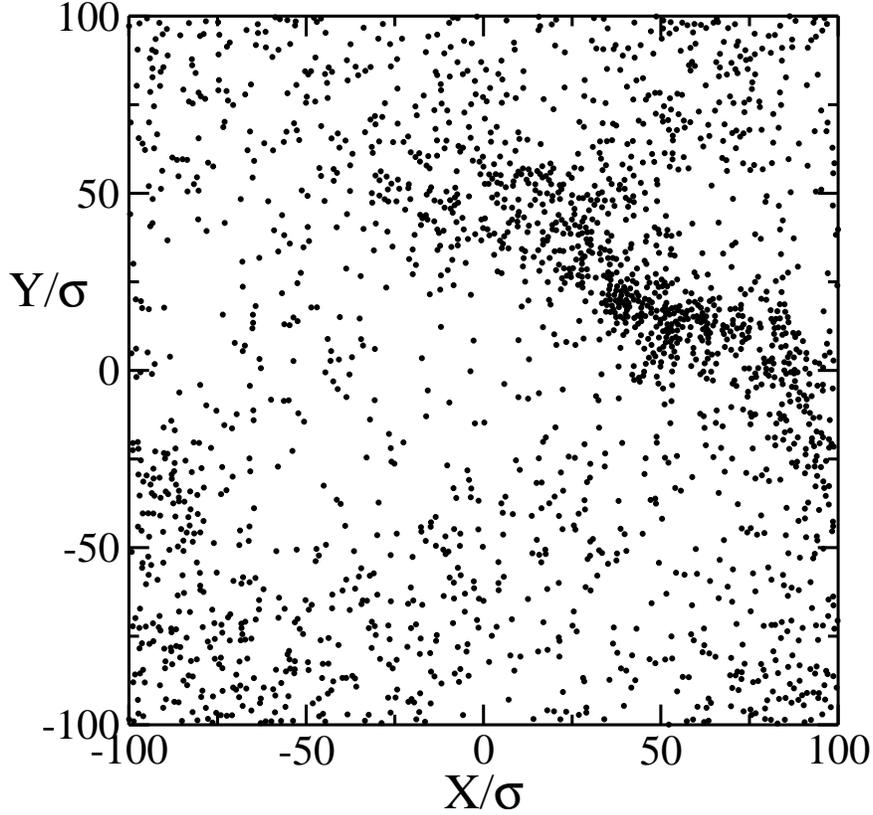}
  \caption{Snapshot of the position of the particles at a time in which the fluctuation of  the Fourier component of the transversal velocity field is large. The system and its phase space trajectory  is the same as in Figs. \ref{fig5} and \ref{fig6}.  The time is $\tau=22.5$, measured in average accumulated number of collisions per particle}
  \label{fig7}
\end{figure}

\begin{figure}
 \includegraphics[width=.7\textwidth]{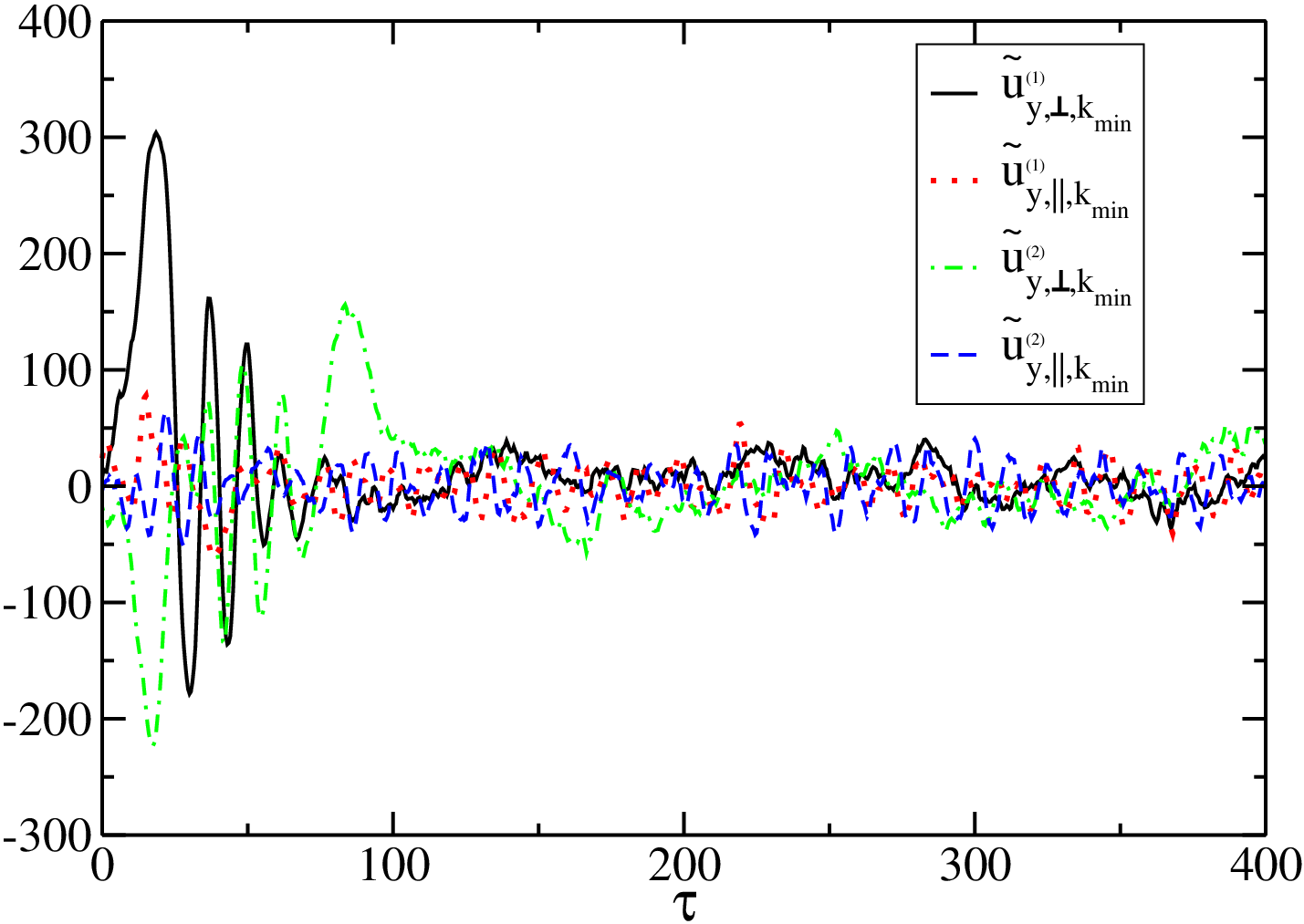}
  \caption{(Color online)  MD simulation results for time evolution of the dimensionless transversal and longitudinal  components of an spontaneous fluctuation of the $y$ component of the velocity field. The wavenumber $k=k_{min}$ is the minimum compatible with the imposed periodic boundary conditions.  Time $\tau$ is the accumulated number of collisions per particle. The parameters of the system are the same as in Fig. \protect{\ref{fig5}}. }
  \label{fig8}
\end{figure}

Similar results have been obtained along other trajectories of the system, i.e. starting from different initial conditions, and also using different number of particles, but the same density. The only relevant difference showing up is the time for which the spontaneous fluctuations of the transversal velocity field take place and also the component ($u_{x}$ or $u_{y}$) that is involved in the initial increase of the fluctuation. Note that the wavevector only can take its minimum value when it is oriented along one of these directions due to the periodic boundary conditions. It is worth to comment what happens with the fluctuations of the velocity field in the direction perpendicular to the one showing a non oscillatory increase of the fluctuations. The simulations show that they oscillate in time  similarly to e.g. the density field. This is illustrated in Fig. \ref{fig8}  for the same system and the same trajectory as considered in Figs, \ref{fig5}-\ref{fig7}.  Once again this is a manifestation of the nonlinear character of the effects we are discussing since in the linear regime the transversal velocity does not exhibit an oscillatory behavior.  

In summary, the MD simulations carried out show a behavior of the fluctuations of the hydrodynamic fields that is consistent with the predictions obtained from the linearized hydrodynamic equations. It is observed that the amplitude of the fluctuations of the transversal velocity field can grow monotonically in time, as predicted, until reaching values for which nonlinear effects are relevant.  The formation of density clusters is observed. Of course, although the linear analysis is able to predict the transitory growth of the transversal mode, it can not explain the formation of density clusters or their spatial structure. These nonlinear effects drive later on the system to the time-dependent homogeneous state, rendering it stable, and consistently, the density clusters disappear.

\section{Conclusion}
\label{s5}
The objective here has been to investigate the stability of the time-dependent homogeneous hydrodynamic state of a model for a confined quasi-two-dimensional granular gas. The main  motivation is to investigate whether the hydrodynamic equations derived for the model are able to predict the bimodal regime that is observed in experiments \cite{OyU98,KyT15}, or at least the existence of some hydrodynamic instability compatible with its existence. Due to the intrinsic nonequilibrium character of granular gases, the  hydrodynamic equations of the model have the peculiarity of exhibiting a homogenous regime. This makes it possible to study the linear stability not only of the steady state eventually reached by the system, but also of the complete hydrodynamic homogeneous trajectory. Showing how this analysis can be carried out in a  particular case is another aim of the present work.
This implies the linearization of the Navier-Stokes equations around a time-dependent state, and the resulting equations happen to have time-dependent coefficients with a dependence that can not be scaled out simply. The consequence is that the equations must be solved numerically.

The analysis indicates a deep difference between homogeneous processes in which the system is heating and those in which it is cooling. While the former are always stable, the latter can in principle be linearly unstable when the time-dependent cooling rate is large enough. This is because the initial value of the eigenvalue associated to the shear mode can be positive indicating a growth of the perturbation of the traversal velocity, at least for a time period after the spontaneous perturbation. To determine what actually happens in the system, molecular dynamics of the model at low density have been performed. The picture emerging from them is that spontaneous fluctuations of the transversal velocity field with an amplitude increasing in time, can actually happen. Moreover they can 
grown enough as to be necessary to consider contributions of terms that are nonlinear in the deviations of the hydrodynamic fields. In particular, clustering of particles are formed. Nevertheless, these nonlinear contributions force the fluctuations to decay to zero for larger times, i.e. the homogeneous time dependent reference state is stable.

The conclusion is that the hydrodynamic equations of the system do not present any instability for homogeneous states, including as a particular case the steady state. Consequently, it is not clear how it could describe the bimodal regime seen in experiments, and some modification of the model seems necessary. An appealing macroscopic approach to this issue has been presented in Ref. \cite{KyA11} and it seems worth trying to understand it on a more fundamental basis level, starting from a particle dynamics description. To put the above comments in a proper context, it is appropriate to emphasize that they do not refer to the solid-fluid phase transition presented by elastic hard spheres or disks. This transition is also present in the model, although slightly modified. Identify the details of the change deserves much more work and it  is now in progress.

\section{Acknowledgements}

This research was supported by the Ministerio de Econom\'{\i}a y Competitividad  (Spain) through Grant No. FIS2014-53808-P (partially financed by FEDER funds).

\appendix*

\section{The hydrodynamic coefficients}

Here the expressions of the coefficients appearing in the constitutive relations (\ref{2.6})-(\ref{2.8}) leading to the hydrodynamic Navier-Stokes equations \cite{BBMyG15},  are reproduced in the units used in the main text. Define dimensionless cooling rate and Navier-Stokes transport coefficients transport coefficients by
\begin{equation}
\label{ap.1}
\overline{\zeta}^{(0)}  \equiv \frac{\zeta^{(0)} \ell }{ v_{0}}\, ,
\end{equation}
\begin{equation}
\label{ap.2}
\overline{\eta} \equiv \frac{\eta }{m n_{H}\ell v_{0}}\, ,
\end{equation}
\begin{equation}
\label{ap.3}
\overline{\kappa} \equiv \frac{\kappa}{n_{H} \ell v_{0}}\, ,
\end{equation}
\begin{equation}
\label{ap.4}
\overline{\mu}  \equiv \frac{\mu}{T \ell v_{0}}\, .
\end{equation}
The expression of the reduced cooling rate    is
\begin{equation}
\label{ap.5}     
  \overline{\zeta}^{(0)}= (2 \pi )^{12/2}\left[\frac{1-\alpha^2}{2}\left(1+\frac{3a_2}{16}\right)-\alpha\left(\frac{\pi}{2}\right)^{1/2}\Delta^*-\left(1-\frac{a_2}{16}\right)\Delta^{*2}\right],
 \end{equation} 
while the transport coefficients are given by the normal solutions, in the sense of being independent from the initial conditions,  of the differential equations
\begin{equation}
\label{ap.6}
\frac{\partial \overline{\eta}}{\partial s}+\left(\bar{\nu}_{\eta}-\frac{\bar{\zeta}^{(0)}}{2}\right)\overline{\eta}=\frac{1}{2},
\end{equation}
\begin{equation}
\label{ap.7}
\frac{\partial \overline{\kappa}}{\partial s}+\left[\bar{\nu}_{\kappa}+\frac{\Delta^*}{2}\frac{\partial \bar{\zeta}^{(0)}}{\partial \Delta^*}-2\bar{\zeta}^{(0)}\right]\overline{\kappa}=1+2a_2-\frac{\Delta^*}{2}\frac{\partial a_2}{\partial \Delta^*},
\end{equation}
\begin{equation}
\label{ap.8}
\frac{\partial \overline{\mu}}{\partial s}+(\bar{\nu}_{\mu}-\frac{3}{2}\bar{\zeta}^{(0)})\overline{\mu}-\bar{\zeta}^{(0)}\overline{\kappa}=a_2.
\end{equation}
In the above expressions, $a_{2}$ is the coefficient of the first Sonine correction to the Gaussian. It obeys the differential equation
\begin{equation}
\label{ap.9}
\frac{\partial a_2}{\partial s}= \frac{\sqrt{2\pi}}{8 \Delta^*}\left\{\left(B_{0}(\Delta^*)+4 A_{0}(\Delta^*)\right)+\left[B_{1}(\Delta^*)+4( A_{1}(\Delta^*)+A_{0}(\Delta^*))\right]a_2
\right\},
\end{equation}
with the several coefficients appearing in it being given by 
\begin{equation}
\label{ap.10}
A_{0} (\alpha, \Delta^{*})= 4 \left[ \frac{1- \alpha^{2}}{2} - \left( \frac{\pi}{2} \right)^{1/2}  \alpha \Delta^{*} - \Delta^{*2} \right],
\end{equation}
\begin{equation}
\label{ap.11}
A_{1} (\alpha, \Delta^{*}) = \frac{1}{4} \left[ \frac{3(1- \alpha^{2})}{2}\, + \Delta^{*2} \right],
\end{equation}
\begin{eqnarray}
\label{ap.12}
B_{0}(\alpha, \Delta^{*}) &=& (2 \pi )^{1/2} \left(5+3 \alpha^{2} +4 \Delta^{*2}\right) \alpha \Delta^{*}-3+4 \Delta^{*4} + \alpha^{2}+ 2 \alpha^{4}  \nonumber  \\
&& -4 \left( 1-\alpha^{2}-2 \Delta^{*2} \right)+ 2 \Delta^{*2} \left( 1+6 \alpha^{2} \right),
\end{eqnarray}
\begin{eqnarray}
\label{ap.13}
B_{1}(\alpha,\Delta^{*})& = & \left( \frac{\pi}{2} \right)^{1/2} \left[ 2 -4(1-\alpha)+7 \alpha+3 \alpha^{3} \right] \Delta^{*}- \frac{1}{16} \left\{ 85 + 4 \Delta^{*4}-18 (3+2 \alpha^{2}) \Delta^{*2} \right.  \nonumber \\
&& \left. - \left( 32 +87 \alpha+ 30 \alpha^{3} \right) \alpha -4 \left[ 6 \Delta^{*2}-(1+\alpha) (31-15 \alpha) \right] \right\}.
\end{eqnarray}

Finally, the Euler transport coefficient in Eq. (\ref{2.8}) has the form
\begin{equation}
\label{ap.14}
\zeta_1=\bar{\zeta}_1 b_2,
\end{equation}
where
\begin{equation}
\label{ap.15}
\bar{\zeta}_1=\frac{\pi^{1/2}}{2^{\frac{25}{2}}}[96+9a_2-3\alpha^2(32+3a_2)+\Delta^{*2}(64+30a_2)]
\end{equation}
and $b_{2}$ is the solution of the differential equation
\begin{equation}
\label{ap.16}
\frac{\partial b_2}{\partial s}-\frac{1}{2}\left[3\bar{\zeta}^{(0)}+\chi+8\bar{\zeta}_1(a_2+1)-2\bar{\zeta}_1\Delta^*\frac{\partial a_2}{\partial \Delta^*}\right]b_2=-\Delta^*\frac{\partial a_2}{\partial \Delta^*} \, ,
\end{equation}
with
\begin{eqnarray}
\label{ap.17}
\chi &=& \frac{\pi^{1/2}}{2^{11}}\left( \sqrt{2}\ \left\{ 30\alpha^4(32-a_2)-5(544+7a_2)-4\Delta^{*2}(32+15a_2) \right. \right. \nonumber \\
&& -64\alpha(16+a_2)-4(992+17a_2)+3\alpha^2[928+43a_2+12\Delta^{*2})(32+3a_2) \nonumber \\
&& \left. 
+20(32-a_2)]+6\Delta^{*2}(288-45a_2+128+12a_2)\right\} \nonumber \\
&& \left. +512\sqrt{\pi}\Delta^*[2+7\alpha+3\alpha^3-4(1-\alpha)] \right).
\end{eqnarray}

\end{document}